\documentclass[aps,twocolumn,showpacs]{revtex4}
\usepackage{amsmath}
\usepackage{epsfig}
\usepackage{booktabs}

\begin{document}
	
\title{Prediction of charmed-bottom pentaquarks in quark model}
\author{Ye Yan$^1$}\email{221001005@njnu.edu.cn}
\author{Yuheng Wu$^1$}\email[E-mail: ]{191002007@njnu.edu.cn}
\author{Hongxia Huang$^1$}\email{hxhuang@njnu.edu.cn(Corresponding author)}
\author{Jialun Ping$^1$}\email{jlping@njnu.edu.cn(Corresponding author)}
\author{Xinmei Zhu$^2$}\email[E-mail: ]{xmzhu@yzu.edu.cn}
\affiliation{$^1$Department of Physics, Nanjing Normal University, Nanjing, Jiangsu 210097, P. R. China}
\affiliation{$^2$Department of Physics, Yangzhou University, Yangzhou 225009, P. R. China}

\begin{abstract}
Inspired by the fully heavy tetraquark states reported by the LHCb, ATLAS and CMS Collaborations, we perform a systemical investigation of the low-lying fully heavy pentaquark systems composed of charm and bottom quarks (anti-quark) in the chiral quark model.
With the help of the channel-coupling, we obtain several fully heavy pentaquark candidates, which are $cccc\bar{b}$ and $bbbb\bar{c}$ systems with $J^P = 1/2^-$ and $3/2^-$, $cccb\bar{c}$, $bbbc\bar{b}$, $cccb\bar{b}$ and $bbbc\bar{c}$ systems with $J^P = 5/2^-$. The binding energies of these states are all below 10 MeV and the root
mean square (RMS) are around 1.8 fm, which indicates that these states are likely to be molecular states. These predicted exotic states may provide new ideas for experimental searches and we expect more experimental and theoretical researches to study and understand the fully heavy states in future.
\end{abstract}
	
	
\maketitle

\setcounter{totalnumber}{5}
	
\section{\label{sec:introduction}Introduction}

Along with baryons and mesons, multi-quark states were also proposed by Gell-Mann and Zweig in the early stage of quark models in 1964~\cite{Gell-Mann:1964ewy,Zweig:1964ruk}.
In recent years, great progress has been made on studying  multi-quark states composed of heavy quarks, since such states with very large energy can be accessed experimentally and easily distinguished from other states, among which the fully heavy tetraquark states $(QQ\bar{Q}\bar{Q}, Q=c,b)$ attracted extensive attention.

In terms of experiment, the CMS collaboration observed pair production of $\Upsilon(1S)$ mesons in proton-proton collisions at $\sqrt{s}=$8 TeV~\cite{CMS:2016liw} and the existence of a fully bottom tetraquark $bb\bar{b}\bar{b}$ in the $\Upsilon(1S) \ell^+ \ell^-$ final states with a mass around 18.4 GeV and a global significance of 3.6 $\sigma$ was claimed~\cite{CMSb,Yi:2018fxo}.
In addition, an evidence of a significance peak around 18.12 GeV was reported by the ANDY Collaboration in Cu + Au collisions at the Relativistic Heavy Ion Collider~\cite{ANDY:2019bfn}.
However, some other experiments searched for the $bb\bar{b}\bar{b}$ tetraquark but no observation was made~\cite{LHCb:2018uwm,CMS:2020qwa}.
In 2020, the LHCb collaboration reported their result on the observation of the fully charmed state $cc\bar{c}\bar{c}$.
A narrow structure $X(6900)$, matching the lineshape of a resonance and a broad structure next to the $J/\psi J/\psi$ mass threshold was obtained~\cite{Aaij:2020fnh}.
At the recent ICHEP 2022 conference, the ATLAS collaboration reported on the di-charmonium resonance search in the four-muon final state with ATLAS 13 TeV data~\cite{ATLAS,ATLAS:2023bft}.
Significant excess observed in the di-$J/\psi$ invariant mass spectrum, and a broad structure at lower mass along with a resonance around 6.9 GeV was observed.
They also observed excess at 6.9 GeV and 7.2 GeV in the $J/\psi + \psi (2S)$ invariant mass spectrum.
The existence of $X(6900)$ was also confirmed by the CMS collaboration at the ICHEP 2022 conference~\cite{CMSs}.
Moreover, the signals of two new peaking structures was observed, which are $X(6600)$ with $6552 \pm 10 \pm 12$ MeV and $X(7200)$ with $7287 \pm 19 \pm 5$ MeV.
These experimental breakthroughs have attracted a lot of attention and inspired a lot of enthusiasm for the theoretical researches on fully heavy states.

Theoretical work in regard to multi-quark states composed of heavy quarks began in 1976.
In ref~\cite{Iwasaki:1975pv}, Iwasaki predicted a fully charmed resonance state at about 6.2 GeV based on a model calculation.
Since then, theoretical studies of the fully heavy states have continued to be carried out and became prosperous since the LHCb and CMS collaborations reported relevant experiments.
The fully heavy tetraquark states have been investigated in the framework of lattice quantum chromodynamics(QCD) method~\cite{Hughes:2017xie}, the QCD sum rules~\cite{Chen:2016jxd, Zhang:2020xtb, Yang:2020wkh, Wan:2020fsk, Wang:2021taf,Wang:2021mma, Yang:2021zrc, Albuquerque:2020hio, Wu:2022qwd, Chen:2022mcr, Wang:2022xja, ZGWang}, relativistic quark model~\cite{Lu:2020cns, Faustov:2021hjs, Faustov:2022mvs, Faustov:2020qfm}, various non-relativistic quark model~\cite{Chao:1980dv, Heller:1985cb, Lloyd:2003yc, Berezhnoy:2011xn, Wu:2016vtq, Richard:2017vry, Esposito:2018cwh, Liu:2019zuc, Wang:2019rdo, Debastiani:2017msn, Jin:2020jfc, Zhao:2020zjh, Lundhammar, Weng:2020jao, Gordillo:2020sgc, Yang:2021hrb, Mutuk:2021hmi, Tiwari:2021iqu, Wang:2021kfv, Liu:2021rtn, Kuang:2022vdy, Wang:2022yes, Zhang:2022qtp, Majarshin:2021hex, An:2022qpt, Hu:2022zdh}, QCD string model ~\cite{Nefediev:2021pww}, flux-tube model~\cite{Bai:2016int}, holography inspired stringy hadron model~\cite{Sonnenschein}, dynamical diquark model~\cite{Giron:2020wpx, Mutuk:2022nkw}, string junction picture~\cite{Karliner:2016zzc, Karliner:2020dta} and Bethe-Salpeter wave function~\cite{Zhu:2020xni, Ke:2021iyh, Li:2021ygk}.
There are also studies investigated the production mechanisms~\cite{Wang:2020gmd, Zhao:2020nwy, Huang:2021vtb, Liang:2021fzr, Dong:2020hxe, Dong:2020nwy, Wang:2020wrp, Cao:2020gul, Goncalves:2021ytq, Feng:2020qee, Maciula:2020wri, Wang:2020tpt, Feng:2020riv, Gong:2020bmg, Niu:2022vqp, Ma:2020kwb, Zhu:2020snb} and the decay properties ~\cite{Li:2019uch, Chen:2020xwe, Becchi:2020uvq, Guo:2020pvt, Chen:2022sbf} of the fully heavy tetraquark states.

On the base of heavy-antiquark-diquark symmetry (HADS)~\cite{Savage:1990di}, which states that two heavy quarks within a doubly heavy baryon behave approximately as a heavy antiquark. If one replaces the anti-charm quark with the charmed di-quark, fully charmed pentaquark and dibaryon configurations can be obtained naturally.
Therefore, the fully heavy pentaquark and dibaryon are quite valuable to study.

The fully heavy dibaryons were also studied in the framework of the the lattice QCD method~\cite{Junnarkar:2019equ, Lyu:2021qsh, Mathur:2022nez}, the QCD sum rules~\cite{Wang:2022jvk}, one-boson exchange model~\cite{Liu:2021pdu}, non-relativistic constituent quark models~\cite{Richard:2020zxb, Huang:2020bmb, Alcaraz-Pelegrina:2022fsi, Lu:2022myk}, potential model~\cite{Richard:2021jgp}, extended chromomagnetic model~\cite{Weng:2022ohh}.
According to the results in Refs.~\cite{Junnarkar:2019equ, Lyu:2021qsh, Mathur:2022nez, Wang:2022jvk, Liu:2021pdu, Huang:2020bmb,Alcaraz-Pelegrina:2022fsi}, $\Omega_{ccc}\Omega_{ccc}$ and $\Omega_{bbb}\Omega_{bbb}$ are possible to form stable bound states.
However, different opinions can be found in Refs.~\cite{Richard:2020zxb,Lu:2022myk, Richard:2021jgp, Weng:2022ohh}.
In their calculations, the energies of $\Omega_{ccc}\Omega_{ccc}$ and $\Omega_{bbb}\Omega_{bbb}$ are above the corresponding thresholds, respectively.

Accordingly, it is timely to study the existence of the fully heavy pentaquarks.
The energies of $cccc\bar{c}$ and $bbbb\bar{b}$ bound states with $J^P=\frac{1}{2}^-$ and $J^P=\frac{3}{2}^-$ have been presented in Refs.~\cite{Zhang:2020vpz,Wang:2021xao} via QCD sum rules approach.
Besdies, systematic studies on the mass spectra of the $S-$wave fully heavy pentaquarks $QQQQ\bar{Q}$ have also been performed in the framework of the chromomagnetic interaction (CMI) model~\cite{An:2020jix}, constituent quark model with variational method~\cite{An:2022fvs}, a lattice-QCD inspired quark model with complex scaling method~\cite{Yang:2022bfu}, quark models with resonating group method~\cite{Yan:2021glh} and MIT bag model~\cite{Zhang:2023hmg}.
Several conclusions concerning resonance states can been found in the Refs.~\cite{An:2020jix, Yang:2022bfu}.
In Refs.~\cite{An:2020jix, An:2022fvs,Zhang:2023hmg}, the author systematicly investigated not only fully charmed (bottom) pentaquarks, but also all possible charmed-bottom pentaquarks with color singlet-singlet strcture ($\bf{1}$$ _{[QQQ]} \bf{\otimes} \bf{1}$$_{[Q \bar{Q}]}$).
In Ref.~\cite{Yang:2022bfu}, the author take into consideration different types of color ($\bf{1}$$ _{[QQQ]} \bf{\otimes} \bf{1}$$_{[Q \bar{Q}]}$ and $\bf{8}$$ _{[QQQ]} \bf{\otimes} \bf{8}$$_{[Q \bar{Q}]}$) and spatial structures to investigate $cccc\bar{c}$ and $bbbb\bar{b}$ states.

The quantum chromodynamics is the underlying theory of the strong interaction.
However, it is difficult to study the structure of the hadrons and the hadron-hadron interaction directly because of the nonperturbative properties of QCD in the low energy region.
Although lattice QCD has made impressive progresses on nucleon-nucleon interactions and multiquark systems~\cite{Ishii:2006ec,Alexandrou:2001ip,Okiharu:2004wy}, the QCD-inspired quark model is still the main approach to study the hadron-hadron interactions and multiquark states.
A common approach is the chiral quark model (ChQM)~\cite{Salamanca1}, in which the constituent quarks interact with each other through colorless Goldstone bosons exchange in addition to the colorful one-gluon-exchange and confinement, and the chiral partner $\sigma$ meson-exchange is introduced to obtain the immediate-range attraction of hadron-hadron interaction.
As for fully heavy pentaquark systems, since there lacks a light-meson exchange mechanism, we mainly consider the one-gluon-exchange and confinement terms.
	
Our purpose is to find out if there exists any fully heavy pentaquark states by investigating the possible charmed-bottom pentaquark systems.
This paper is organized as follows.
After introduction, we briefly introduce the quark model used in section II.
The way we construct the wave functions is also presented.
Then, the numerical results and discussions are presented in Section III.
Finally, the paper ends with summary in Section IV.
	
\section{THEORETICAL FRAMEWORK}
Herein, ChQM is employed to investigate the properties of pentaquark systems, and the channel coupling effect is considered.
Both color singlet-singlet structure (color singlet channel) and color octet-octet structure (hidden-color channel) are taken into consideration.
In this sector, we will introduce this model and the way of constructing wave functions.
	
\subsection{Chiral quark model (ChQM)}
The Salamanca version was chosen as the representative of the chiral quark models, because the Salamanca group's work covers the hadron spectra,
nucleon-nucleon interaction, and multiquark states. The model details can be found in Ref.~\cite{Salamanca1}.
Here, we mainly present the salient features of the model.
The general form of the five body complex Hamiltonian is given by
	\begin{align}
		H=&\sum_{i=1}^5\left(m_i+\frac{\boldsymbol{p}_{i}^{2}}{2m_i}\right)-T_{CM} +\sum_{j>i=1}^5 V(\boldsymbol{r}_{ij})
	\end{align}
where $m_i$ is the quark mass, $\boldsymbol{p}_{i}$ is the momentum of the quark, and $T_{CM}$ is the center-of-mass kinetic energy.
The dynamics of the pentaquark system is driven by a two-body potential
	\begin{align}
		V(\boldsymbol{r}_{ij})= & V_{CON}(\boldsymbol{r}_{ij})+V_{OGE}(\boldsymbol{r}_{ij})
	\end{align}
The most relevant features of QCD at its low energy regime: color confinement ($V_{CON}$) and perturbative one-gluon exchange interaction ($V_{OGE}$) have been taken into consideraton.
For the fully-heavy systems, there is no $\sigma-$exchange or the Goldstone boson exchange.

Here, a phenomenological quadratic form confinement potential($V_{CON}$) is used as
\begin{align}
	V_{CON}(\boldsymbol{r}_{ij}) = & -a_{c}\boldsymbol{\lambda}_{i}^{c} \cdot \boldsymbol{\lambda}_{j}^{c}\left[  f(\boldsymbol{r}_{ij})+V_{0}\right]
\end{align}

In the present work, we mainly focus on the low-lying negative parity $QQQQ\bar{Q}$ pentaquark systems of S-wave, so the spin-orbit and tensor interactions are not included.
The one-gluon exchange potential ($V_{OGE}$), which includes coulomb and color-magnetic interactions, is written as
	\begin{align}
		V_{OGE}(\boldsymbol{r}_{ij})= &\frac{1}{4}\alpha_{s} \boldsymbol{\lambda}_{i}^{c} \cdot \boldsymbol{\lambda}_{j}^{c} \nonumber \\
		&\cdot \left[\frac{1}{r_{i j}}-\frac{\pi}{2} \delta\left(\mathbf{r}_{i j}\right)\left(\frac{1}{m_{i}^{2}}+\frac{1}{m_{j}^{2}}+\frac{4 \boldsymbol{\sigma}_{i} \cdot \boldsymbol{\sigma}_{j}}{3 m_{i} m_{j}}\right)\right]    \label{Voge}
	\end{align}
where $\boldsymbol{\sigma}$ is the Pauli matrices and $\alpha_{s}$ is the quark-gluon coupling constant. In order to cover the wide energy range from light to strange to heavy quarks, an effective scale-dependent quark-gluon coupling $\alpha_{s}(\mu)$ between quarks was introduced ~\cite{Vijande:2004he}:
	\begin{align}
		\alpha_{s}(\mu) =& \frac{\alpha_{0}}{\ln \left(\frac{\mu^{2}+\mu_{0}^{2}}{\Lambda_{0}^{2}}\right)}
	\end{align}

The other symbols in the above expressions have their usual meanings.
All parameters, which are fixed by fitting to the masses of baryons with light flavors and heavy flavors, are taken from our previous work Ref.~\cite{Huang:2019esu, Huang:2020bmb}.

\subsection{Wave functions}
The resonating group method (RGM)~\cite{RGM1,RGM} and generating coordinates method~\cite{GCM1,GCM2} are used to carry out a dynamical calculation.
The main feature of the RGM for two-cluster systems is that it assumes that two clusters are frozen inside, and only considers the relative motion between the two clusters.
So the conventional ansatz for the two-cluster wavefunctions is
\begin{equation}
	\psi_{5q} = {\cal A }\left[[\phi_{B_{1}}\phi_{B_{2}}]^{[\sigma]IS}\otimes\chi_{L}(\boldsymbol{R})\right]^{J} \label{5q}
\end{equation}
where the symbol ${\cal A }$ is the anti-symmetrization operator. $[\sigma]=[222]$ gives the total color symmetry and all other symbols have their usual meanings.
 $\phi_{B_{1}}$ and $\phi_{B_{2}}$ are the $q^{3}$ and $\bar{q} q$ cluster wavefunctions, respectively.
 From the variational principle, after variation with respect to the relative motion wave function $\chi(\boldsymbol{\mathbf{R}})=\sum_{L}\chi_{L}(\boldsymbol{\mathbf{R}})$, one obtains the RGM equation:
\begin{equation}
	\int H(\boldsymbol{\mathbf{R}},\boldsymbol{\mathbf{R'}})\chi(\boldsymbol{\mathbf{R'}})d\boldsymbol{\mathbf{R'}}=E\int N(\boldsymbol{\mathbf{R}},\boldsymbol{\mathbf{R'}})\chi(\boldsymbol{\mathbf{R'}})d\boldsymbol{\mathbf{R'}}  \label{RGM eq}
\end{equation}
where $H(\boldsymbol{\mathbf{R}},\boldsymbol{\mathbf{R'}})$ and $N(\boldsymbol{\mathbf{R}},\boldsymbol{\mathbf{R'}})$ are Hamiltonian and norm kernels.
By solving the RGM equation, we can get the energies $E$ and the wave functions.
In fact, it is not convenient to work with the RGM expressions.
Then, we expand the relative motion wave function $\chi(\boldsymbol{\mathbf{R}})$ by using a set of gaussians with different centers
\begin{align}
	\chi_{L}(\boldsymbol{R}) =& \frac{1}{\sqrt{4 \pi}}\left(\frac{6}{5 \pi b^{2}}\right)^{3 / 4} \sum_{i = 1}^{n} C_{i} \\ \nonumber    &\cdot\int \exp \left[-\frac{3}{5 b^{2}}\left(\boldsymbol{R}-\boldsymbol{S}_{i}\right)^{2}\right] Y^{L}\left(\hat{\boldsymbol{S}}_{i}\right) d \hat{\boldsymbol{S}}_{i}
\end{align}
where $L$ is the orbital angular momentum between two clusters, and $\boldsymbol {S_{i}}$, $i=1,2,...,n$ are the generator coordinates, which are introduced to expand the relative motion wavefunction. By including the center of mass motion:
\begin{equation}
	\phi_{C} (\boldsymbol{R}_{C}) = (\frac{5}{\pi b^{2}})^{3/4}e^{-\frac{5\boldsymbol{R}^{2}_{C}}{2b^{2}}}
\end{equation}
the ansatz Eq.(\ref{5q}) can be rewritten as
\begin{align}
	\psi_{5 q} =& \mathcal{A} \sum_{i=1}^{n} C_{i} \int \frac{d \hat{\boldsymbol{S}}_{i}}{\sqrt{4 \pi}} \prod_{\alpha=1}^{3} \phi_{\alpha}\left(\boldsymbol{S}_{i}\right) \prod_{\beta=4}^{5} \phi_{\beta}\left(-\boldsymbol{S}_{i}\right) \nonumber \\
	& \cdot\left[\left[\chi_{I_{1} S_{1}}\left(B_{1}\right) \chi_{I_{2} S_{2}}\left(B_{2}\right)\right]^{I S} Y_{L M}\left(\hat{\boldsymbol{S}}_{i}\right)\right]^{J} \nonumber \\
	& \cdot\left[\chi_{c}\left(B_{1}\right) \chi_{c}\left(B_{2}\right)\right]^{[c]} \label{5q2}
\end{align}
where $\chi_{I_{1}S_{1}}$ and $\chi_{I_{2}S_{2}}$ are the product of the flavor and spin wave functions, and $\chi_{c}$ is the color wave function. These will be shown in detail later.  $\phi_{\alpha}(\boldsymbol{S}_{i})$ and $\phi_{\beta}(-\boldsymbol{S}_{i})$ are the single-particle orbital wavefunctions with different reference centers:
\begin{align}
	\phi_{\alpha}\left(\boldsymbol{S}_{i}\right) & = \left(\frac{1}{\pi b^{2}}\right)^{3 / 4} e^{-\frac{1}{2 b^{2}}\left(r_{\alpha}-\frac{2}{5} \boldsymbol{S}_{i}\right)^{2}} \\ \nonumber
	\phi_{\beta}\left(\boldsymbol{-S}_{i}\right) & = \left(\frac{1}{\pi b^{2}}\right)^{3 / 4} e^{-\frac{1}{2 b^{2}}\left(r_{\beta}+\frac{3}{5} \boldsymbol{S}_{i}\right)^{2}}
\end{align}
With the reformulated ansatz Eq.(\ref{5q2}), the RGM Eq.(\ref{RGM eq}) becomes an algebraic eigenvalue equation:
\begin{equation}
	\sum_{j} C_{j}H_{i,j}= E \sum_{j} C_{j}N_{i,j}
\end{equation}
where $H_{i,j}$ and $N_{i,j}$ are the Hamiltonian matrix elements and overlaps, respectively.
By solving the generalized eigen problem, we can obtain the energy and the corresponding wavefunctions of the pentaquark systems.

For the spin wave function, we first construct the spin wave functions of the $q^{3}$ and $\bar{q}q$ clusters with SU(2) algebra, and then the total spin wave function of the pentaquark system is obtained by coupling the spin wave functions of two clusters together.
The spin wave functions of the $q^{3}$ and $\bar{q}q$ clusters are
\begin{align}
	\chi_{\frac{3}{2}, \frac{3}{2}}^{\sigma}(3) &=  
	\big|
	\begin{tabular}{|c|c|c|}
		\hline
		1 & 2 & 3 \\
		\hline
	\end{tabular} \,\,
	\begin{tabular}{|c|c|c|}
		\hline
		$\alpha$ & $\alpha$ & $\alpha$ \\
		\hline
	\end{tabular}
	\,\big > = \alpha \alpha \alpha
	\nonumber \\
	\chi_{\frac{3}{2}, \frac{1}{2}}^{\sigma}(3) & = 
	\big|
	\begin{tabular}{|c|c|c|}
		\hline
		1 & 2 & 3 \\
		\hline
	\end{tabular} \,\,
	\begin{tabular}{|c|c|c|}
		\hline
		$\alpha$ & $\alpha$ & $\beta$ \\
		\hline
	\end{tabular}
	\,\big > =  \frac{1}{\sqrt{3}}(\alpha \alpha \beta+\alpha \beta \alpha+\beta \alpha \alpha)
	\nonumber \\
	\chi_{\frac{3}{2}, -\frac{1}{2}}^{\sigma}(3) & =  
	\big|
	\begin{tabular}{|c|c|c|}
		\hline
		1 & 2 & 3 \\
		\hline
	\end{tabular} \,\,
	\begin{tabular}{|c|c|c|}
		\hline
		$\alpha$ & $\beta$ & $\beta$ \\
		\hline
	\end{tabular}
	\,\big > =  \frac{1}{\sqrt{3}}(\alpha \beta \beta+\beta \alpha \beta+\beta \beta \alpha)
	\nonumber \\
	\chi_{\frac{3}{2}, -\frac{3}{2}}^{\sigma}(3) &= 
	\big|
	\begin{tabular}{|c|c|c|}
		\hline
		1 & 2 & 3 \\
		\hline
	\end{tabular} \,\,
	\begin{tabular}{|c|c|c|}
		\hline
		$\beta$ & $\beta$ & $\beta$ \\
		\hline
	\end{tabular}
	\,\big > = \beta \beta \beta
	\nonumber \\
	\chi_{\frac{1}{2}, \frac{1}{2}}^{\sigma 1}(3) & = 
	\,\big|
	\begin{tabular}{|c|c|}
		\hline
		1 & 2 \\
		\hline
		3 \\
		\cline{1-1}
	\end{tabular} \,\,
	\begin{tabular}{|c|c|}
		\hline
		$\alpha$ & $\alpha$ \\
		\hline
		$\beta$ \\
		\cline{1-1}
	\end{tabular}
	\,\big > = \frac{1}{\sqrt{6}}(2 \alpha \alpha \beta-\alpha \beta \alpha-\beta \alpha \alpha)
	\nonumber \\
	\chi_{\frac{1}{2}, \frac{1}{2}}^{\sigma 2}(3) & = 
	\,\big|
	\begin{tabular}{|c|c|}
		\hline
		1 & 3 \\
		\hline
		2 \\
		\cline{1-1}
	\end{tabular} \,\,
	\begin{tabular}{|c|c|}
		\hline
		$\alpha$ & $\alpha$ \\
		\hline
		$\beta$ \\
		\cline{1-1}
	\end{tabular}
	\,\big > = \frac{1}{\sqrt{2}}(\alpha \beta \alpha-\beta \alpha \alpha)
	\nonumber \\
	\chi_{\frac{1}{2},-\frac{1}{2}}^{\sigma1}(3) & = 
	\,\big|
	\begin{tabular}{|c|c|}
		\hline
		1 & 2 \\
		\hline
		3 \\
		\cline{1-1}
	\end{tabular} \,\,
	\begin{tabular}{|c|c|}
		\hline
		$\alpha$ & $\beta$ \\
		\hline
		$\beta$ \\
		\cline{1-1}
	\end{tabular}
	\,\big > = \frac{1}{\sqrt{6}}(\alpha \beta \beta+\beta \alpha \beta-2 \beta \beta \alpha)
	\nonumber \\
	\chi_{\frac{1}{2},-\frac{1}{2}}^{\sigma2}(3) & = 
	\,\big|
	\begin{tabular}{|c|c|}
		\hline
		1 & 3 \\
		\hline
		2 \\
		\cline{1-1}
	\end{tabular} \,\,
	\begin{tabular}{|c|c|}
		\hline
		$\alpha$ & $\beta$ \\
		\hline
		$\beta$ \\
		\cline{1-1}
	\end{tabular}
	\,\big > = \frac{1}{\sqrt{2}}(\alpha \beta \beta-\beta \alpha \beta)
	\nonumber \\
	\chi_{1,1}^{\sigma}(2) &  =
	\,\big|
	\begin{tabular}{|c|c|}
		\hline
		1 & 2 \\
		\hline
	\end{tabular} \,\,
	\begin{tabular}{|c|c|}
		\hline
		$\alpha$ & $\alpha$ \\
		\hline
	\end{tabular}
	\,\big > =  \alpha  \alpha
	\nonumber \\
	\chi_{1,0}^{\sigma}(2) &  =
	\,\big|
	\begin{tabular}{|c|c|}
		\hline
		1 & 2 \\
		\hline
	\end{tabular} \,\,
	\begin{tabular}{|c|c|}
		\hline
		$\alpha$ & $\beta$ \\
		\hline
	\end{tabular}
	\,\big > =  \frac{1}{\sqrt{2}}(\alpha \beta+\beta \alpha)
	\nonumber \\
	\chi_{1,-1}^{\sigma}(2) &  =
	\,\big|
	\begin{tabular}{|c|c|}
		\hline
		1 & 2 \\
		\hline
	\end{tabular} \,\,
	\begin{tabular}{|c|c|}
		\hline
		$\beta$ & $\beta$ \\
		\hline
	\end{tabular}
	\,\big > =  \beta \beta
	\nonumber \\
	\chi_{0,0}^{\sigma}(2) & =  
	\,\big|
	\begin{tabular}{|c|}
		\hline
		1  \\
		\hline
		2 \\
		\hline
	\end{tabular} \,\,
	\begin{tabular}{|c|}
		\hline
		$\alpha$  \\
		\hline
		$\beta$ \\
		\hline
	\end{tabular}
	\,\big > =  \frac{1}{\sqrt{2}}(\alpha \beta-\beta \alpha)
\end{align}
	
For pentaquark system, the total spin quantum number can be 1/2, 3/2 or 5/2, so the wave function of each spin quantum number can be written as follows, according to Clebsch-Gordan (CG) coefficient
	\begin{align}
		\chi_{\frac{5}{2}, \frac{5}{2}}^{\sigma}(5) = & \chi_{\frac{3}{2}, \frac{3}{2}}^{\sigma}(3) \chi_{1,1}^{\sigma}(2) \nonumber\\
		\chi_{\frac{3}{2}, \frac{3}{2}}^{\sigma 1}(5) = & \sqrt{\frac{3}{5}} \chi_{\frac{3}{2}, \frac{3}{2}}^{\sigma}(3) \chi_{1,0}^{\sigma}(2)-\sqrt{\frac{2}{5}} \chi_{\frac{3}{2}, \frac{1}{2}}^{\sigma}(3) \chi_{1,1}^{\sigma}(2)\nonumber \\
        \chi_{\frac{3}{2}, \frac{3}{2}}^{\sigma 2}(5) = & \chi_{\frac{3}{2}, \frac{3}{2}}^{\sigma}(3) \chi_{0,0}^{\sigma}(2)  \nonumber \\
		\chi_{\frac{1}{2}, \frac{1}{2}}^{\sigma 1}(5) = &\chi_{\frac{1}{2}, \frac{1}{2}}^{\sigma}(3) \chi_{0,0}^{\sigma}(2) \nonumber\\
		\chi_{\frac{1}{2}, \frac{1}{2}}^{\sigma 2}(5) = & \frac{1}{\sqrt{6}} \chi_{\frac{3}{2},-\frac{1}{2}}^{\sigma}(3) \chi_{1,1}^{\sigma}(2)-\frac{1}{\sqrt{3}} \chi_{\frac{3}{2}, \frac{1}{2}}^{\sigma}(3) \chi_{1,0}^{\sigma}(2)\nonumber \\
		 &+\frac{1}{\sqrt{2}} \chi_{\frac{3}{2}, \frac{3}{2}}^{\sigma}(3) \chi_{1,-1}^{\sigma}(2)
	\end{align}
		
Similar to constructing spin wave functions, we first write down the flavor wave functions of the $q^{3}$ and $\bar{q} q$ clusters, which are
\begin{align}
	\chi^{f 1}(3) & = ccc  \nonumber  \\
	\chi^{f 2}(3) & = \frac{1}{\sqrt{3}}(ccb+cbc+bcc)  \nonumber  \\
	\chi^{f 3}(3) & = \frac{1}{\sqrt{3}}(bbc+bcb+cbb)  \nonumber  \\
	\chi^{f 4}(3) & = bbb  	\nonumber  \\
	\chi^{f 5}(3) & = \frac{1}{\sqrt{6}}(2ccb-cbc-bcc)  \nonumber  \\
	\chi^{f 6}(3) & = \frac{1}{\sqrt{2}}(cbc-bcc)  \nonumber  \\
	\chi^{f 7}(3) & = \frac{1}{\sqrt{6}}(cbb+bcb-2bbc)  \nonumber  \\
	\chi^{f 8}(3) & = \frac{1}{\sqrt{2}}(cbb-bcb)  \nonumber  \\
	\chi^{f 1}(2) & = \bar{c} c 	\nonumber  \\
	\chi^{f 2}(2) & = \bar{c} b     \nonumber  \\
	\chi^{f 3}(2) & = \bar{b} c     \nonumber  \\
	\chi^{f 4}(2) & = \bar{b} b
\end{align}
As for the flavor degree-of-freedom, since the quark content of the investigated pentaquark system is $QQQQ\bar{Q}$ $(Q = c, b)$, the isospin $I$ of pentaquark states can only be 0.
So the CG coefficient of this part is
\begin{align}
\chi_{0,0}^{f}(5)=\chi_{0,0}^{f}(3)  \chi_{0,0}^{f}(2)
\end{align}

For the color singlet channel, which means two clusters are color singlet-singlet structure, the color wave function can be obtained by $\bf{1}$$ _{[QQQ]} \bf{\otimes} \bf{1}$$_{[Q \bar{Q}]}$
\begin{align}
	\chi^{c 1} =& \frac{1}{\sqrt{6}}(r g b-r b g+g b r-g r b+b r g-b g r) \nonumber \\
	&\cdot\frac{1}{\sqrt{3}}(r \bar{r}+g \bar{g}+b \bar{b})
\end{align}
In addition, it is possible for the pentaquark systems to have hidden-color channel, that is, the baryon and the meson are color octet-octet structure and then become colorless through the coupling of the two clusters( $\bf{8}$$ _{[QQQ]} \bf{\otimes} \bf{8}$$_{[Q \bar{Q}]}$).
The corresponding color wave function of the pentaquark systems can be obtained through the following coupling
\begin{align}
	\big|
	\begin{tabular}{|c|c|}
		\hline
		$r$ & $r$  \\
		\hline
		$g$ & $g$  \\
        \hline
        $b$ & $b$  \\
        \hline
	\end{tabular} \,\big > = \frac{1}{\sqrt{8}} (  ~&
    \big|
    \begin{tabular}{|c|c|}   
    	\hline
    	$r$ & $r$ \\
    	\hline
    	$g$ \\
    	\cline{1-1}
    \end{tabular} \,\big > \otimes
    \big|
    \begin{tabular}{|c|c|}
    	\hline
    	$g$ & $b$ \\
    	\hline
    	$b$ \\
    	\cline{1-1}
    \end{tabular} \,\big > -
    \big|
    \begin{tabular}{|c|c|}   
    	\hline
    	$r$ & $g$ \\
    	\hline
    	$g$ \\
    	\cline{1-1}
    \end{tabular} \,\big > \otimes
    \big|
    \begin{tabular}{|c|c|}
    	\hline
    	$r$ & $b$ \\
    	\hline
    	$b$ \\
    	\cline{1-1}
    \end{tabular} \,\big >
    \nonumber  \\  - ~&
    \big|
    \begin{tabular}{|c|c|}   
    	\hline
    	$r$ & $r$ \\
    	\hline
    	$b$ \\
    	\cline{1-1}
    \end{tabular} \,\big > \otimes
    \big|
    \begin{tabular}{|c|c|}
    	\hline
    	$g$ & $g$ \\
    	\hline
    	$b$ \\
    	\cline{1-1}
    \end{tabular} \,\big > +
    \big|
    \begin{tabular}{|c|c|}   
    	\hline
    	$r$ & $g$ \\
    	\hline
    	$b$ \\
    	\cline{1-1}
    \end{tabular} \,\big > \otimes
    \big|
    \begin{tabular}{|c|c|}
    	\hline
    	$r$ & $g$ \\
    	\hline
    	$b$ \\
    	\cline{1-1}
    \end{tabular} \,\big >
    \nonumber  \\  - ~&
    \big|
    \begin{tabular}{|c|c|}   
    	\hline
    	$g$ & $g$ \\
    	\hline
    	$b$ \\
    	\cline{1-1}
    \end{tabular} \,\big > \otimes
    \big|
    \begin{tabular}{|c|c|}
    	\hline
    	$r$ & $r$ \\
    	\hline
    	$b$ \\
    	\cline{1-1}
    \end{tabular} \,\big > +
    \big|
    \begin{tabular}{|c|c|}   
    	\hline
    	$r$ & $b$ \\
    	\hline
    	$g$ \\
    	\cline{1-1}
    \end{tabular} \,\big > \otimes
    \big|
    \begin{tabular}{|c|c|}
    	\hline
    	$r$ & $b$ \\
    	\hline
    	$g$ \\
    	\cline{1-1}
    \end{tabular} \,\big >
    \nonumber  \\  - ~ &
        \big|
    \begin{tabular}{|c|c|}   
    	\hline
    	$r$ & $b$ \\
    	\hline
    	$b$ \\
    	\cline{1-1}
    \end{tabular} \,\big > \otimes
    \big|
    \begin{tabular}{|c|c|}
    	\hline
    	$r$ & $g$ \\
    	\hline
    	$g$ \\
    	\cline{1-1}
    \end{tabular} \,\big > +
    \big|
    \begin{tabular}{|c|c|}   
    	\hline
    	$g$ & $b$ \\
    	\hline
    	$b$ \\
    	\cline{1-1}
    \end{tabular} \,\big > \otimes
    \big|
    \begin{tabular}{|c|c|}
    	\hline
    	$r$ & $r$ \\
    	\hline
    	$g$ \\
    	\cline{1-1}
    \end{tabular} \,\big >  )
\end{align}
Then, the colorless pentaquark wave functions can be written as
\begin{align}
	\chi^{c 2}= & \frac{1}{\sqrt{48}}(2 r r g-r g r-g r r) \bar{r} b+\frac{1}{\sqrt{48}}(r g g+g r g-2 g g r) \bar{g} b \nonumber \\
	-&\frac{1}{\sqrt{48}}(2 r r b-r b r-b r r) \bar{r} g-{\frac{1}{\sqrt48}}(r b b+b r b-2 b b r) \bar{b} g\nonumber \\
	+& \frac{1}{\sqrt{48}}(2 g g b-g b g-b g g) \bar{g} r+{\frac{1}{\sqrt48}}(g b b+b g b-2 b b g) \bar{b} r \nonumber \\
	+& \frac{1}{\sqrt{192}}(r b g-g b r+b r g-b g r)(2 \bar{b} b-\bar{r} r-\bar{g} g) \nonumber \\
	+& \frac{1}{\sqrt{192}}(2 r g b-r b g+2 g r b-g b r-b r g-b g r)(\bar{r} r-\bar{g} g) \nonumber \\
	\nonumber \\
	\chi^{c 3}  = & \frac{1}{\sqrt{16}}(r g r-g r r) \bar{r} b+\frac{1}{\sqrt{16}}(r g g-g r g) \bar{g} b \nonumber \\
	-&\frac{1}{\sqrt{16}}(r b r-b r r) \bar{r} g-\frac{1}{\sqrt{16}}(r b b-b r b) \bar{b} g \nonumber \\
	+&\frac{1}{\sqrt{16}}(g b g-b g g) \bar{g} r+\frac{1}{\sqrt{16}}(g b b-b g b) \bar{b} r \nonumber \\
	+&\frac{1}{\sqrt{64}}(r b g+g b r-b r g-b g r)(\bar{r} r-\bar{g} g) \nonumber \\
	+&\frac{1}{\sqrt{576}}(2 r g b+r b g-2 g r b-g b r-b r g+b g r) \nonumber  \\
	&  \cdot  (2 \bar{b} b-\bar{g} g-\bar{r} r)
\end{align}

Finally, we can acquire the total wave function by combining the wave functions of the orbital, spin, flavor and color parts together according to the quantum number of the pentaquark systems.
More details on the construction of the wave function can be found in Ref.~\cite{Xia:2021tof}.
		
\section{The results and discussions}
In this work,  we perform a systematical investigation of the $S-$wave fully heavy pentaquark systems $QQQQ\bar{Q}$ ($Q=c,~b$) in the framework of chiral quark model.
The quantum numbers of these systems are $J^P = 1/2^-$, $3/2^-$, and $5/2^-$, respectively.
The effective potentials are calculated to study the interaction between two hadron clusters. The dynamic bound state calculations are carried out to find if there is any bound state. Both color singlet (with color coupling $\bf{1}$$ _{[QQQ]} \bf{\otimes} \bf{1}$$_{[Q \bar{Q}]}$) and hidden color channels (with color coupling $\bf{8}$$ _{[QQQ]} \bf{\otimes} \bf{8}$$_{[Q \bar{Q}]}$), as well as the channel-coupling of all channels are included in the calculation. Moreover, the calculation of root mean square (RMS) is helped to explore the structure of the bound states


All the calculating energy for fully heavy pentaquark systems are listed in Tables \ref{table ccccb}, \ref{table cccbc} and \ref{table ccbbc}, respectively.
The second column, headed with $channel$, denotes the physical contents of each channel; $E_{th}$ and $E_{sc}$ represent the threshold and the energy of each  single channel.
Regarding the results of the channel coupling calculation, $E_{cc1}$ stands for the coupling of only color singlet channels, and $E_{cc2}$ stands for the coupling of both the color singlet channels and the hidden-color channels.
The subscript '$8$' is marked to represent the hidden-color channel.
The last column, headed with $E_B$, stands for the binding energy of the pentaquark state, which is defined as $E_B = E_{cc2} - M_{1} - M_{2}$.
$M_{1}$ and $M_{2}$ are the theoretical masses of the baryon and meson, respectively.

\subsection{$cccc\bar{b}$ and $bbbb\bar{c}$ systems}
\begin{figure*}[htbp]
	\centering
	\includegraphics[width=17cm]{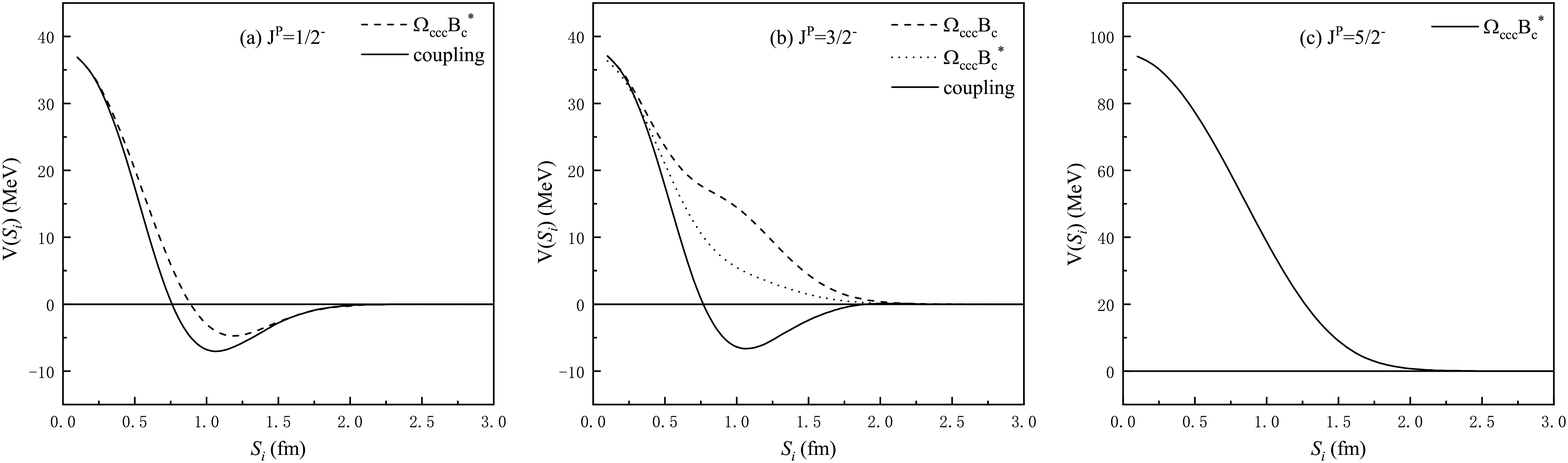}\
	\caption{\label{ccccb} The effective potentials of $cccc\bar{b}$ systems.}
\end{figure*}
Since the attractive potential is necessary for forming bound states, we first calculate the effective potential between the baryon and meson, which is defined as $V(S_{i}) = E(S_{i}) - E(\infty)$, where $E(S_{i})$ is the diagonal matrix element of the Hamiltonian of the system in the generator coordinate.
We present the effective potentials of color singlet channels and channel coupling to investigate the interactions between baryons and mesons.

The effective potentials of $cccc\bar{b}$ systems with different quantum numbers are shown in Fig.~\ref{ccccb}, where 'coupling' stands for the effective potential with coupling of both color singlet and hidden-color channels. The behavior of effective potentials of $bbbb\bar{c}$ systems is similar to $cccc\bar{b}$ systems because of the heavy flavor symmetry.
To save place, we only present the effective potentials of $cccc\bar{b}$ systems.
For the $J^P=1/2^-$ system, one sees that the potential for the channel $\Omega_{ccc} B_{c}^*$ is attractive.
And it is obvious that the attraction becomes deeper after channel-coupling calculation.
For the $J^P=3/2^-$ system, we can see that the effective potentials for the $\Omega_{ccc} B_{c}$ and $\Omega_{ccc} B_{c}^*$ are repulsive.
So it is impossible for these two single channels to form any bound state. However, the effective potential appears attractive after the channel coupling.
This shows that the influence of hidden-color channel and channel coupling is significant in this system.
As for the $J^P=5/2^-$ system, the effective potential is much more repulsive, comparing to the previous two systems.
Moreover, the repulsion increases greatly when the two hadrons $\Omega_{ccc}$ and $B_{c}^*$ get very close.
This is due to the Pauli exclusion principle.
Considering that the four charm quarks in the pentaquark system with $J^P=5/2^-$ are identical particles, they have the same state in the orbital, flavor, and spin wave functions.
But there are only three color states, which are red, green and blue.
Then two quarks must be in the same state, which is prohibited by the Pauli exclusion principle and these four quarks are difficult to stay together.
That's why strong repulsion appears between $\Omega_{ccc}$ and $B_{c}^*$ and the $cccc\bar{b}$ pentaquark state with $J^P=5/2^-$ is difficult to form a bound state.

\begin{table}[htbp]
	\caption{\label{table ccccb}The energies of the $cccc\bar{b}$ and $bbbb\bar{c}$ pentaquark systems (unit: MeV).}
	\begin{tabular}{c c c c c c c} \hline\hline
		
		\multicolumn{7}{c}{$cccc\bar{b}$ systems}  \\ \hline
		$J^P$~  & channel & ~$E_{th}$~ & ~$E_{sc}$~ & ~~$E_{cc1}$~~ & ~~$E_{cc2}$~~& $E_B$  \\ \hline
		
		$\frac{1}{2}^-$ & $\Omega_{ccc} B^*_c$ & ~11777.3~ & ~11778.4~ & ~11778.4~ & ~11776.4~ & $-0.9$  \\
		~ &$\Omega_{ccc8} B_{c8}$& ~ & 11983.0 & ~ & ~\\
		~ &$\Omega_{ccc8} B^*_{c8}$  & ~ & 12065.6 & ~ &  \\ \hline
		
		$\frac{3}{2}^-$&$\Omega_{ccc} B_{c}$ & 11776.5 & 11778.3 & 11778.0 & 11776.0 & $-0.5$ \\
		~ & $\Omega_{ccc} B^*_c$ & 11777.3 & 11778.9 & ~ & ~     \\
		~ & $\Omega_{ccc8} B^*_{c8}$ & ~   & 11940.8 & ~ & ~  \\  \hline
		
		$\frac{5}{2}^- $ & $\Omega_{ccc} B^*_c$ & 11777.3 & 11779.2 & 11779.2 & 11779.2 & ub  \\  \hline\hline

		\multicolumn{7}{c}{$bbbb\bar{c}$ systems}  \\ \hline
		$J^P$~  & channel & ~$E_{th}$~ & ~$E_{sc}$~ & ~~$E_{cc1}$~~ & ~~$E_{cc2}$~~& $E_B$  \\ \hline
		$\frac{1}{2}^-$  & $\Omega_{bbb} B^*_c$  & ~21828.1~ & ~21828.7~ & ~21828.7~ & ~21820.7~ & $-7.4$  \\
		~ &$\Omega_{bbb8} B_{c8}$& ~ & ~22057.8~ & ~ & ~ \\
		~ &$\Omega_{bbb8} B^*_{c8}$  & ~ & 21986.9 & ~ &  \\  \hline
		
		$\frac{3}{2}^-$&$\Omega_{bbb} B_{c}$ & 21827.3 & 21828.3 & 21828.2 & ~21820.3~ & $-7.0$   \\
		~ & $\Omega_{bbb} B^*_c$ & 21828.1 & 21828.9 & ~ & ~   \\
		~ & $\Omega_{bbb8} B^*_{c8}$ & ~   & 21951.0  & ~ &   \\  \hline
		
		$\frac{5}{2}^- $ & $\Omega_{ccc} B^*_c$ & 21828.1 & 21829.1 & ~21829.1~ & ~21829.1~ & ub  \\  \hline\hline
	\end{tabular}
\end{table}
In order to investigate whether there is any bound state, a dynamic calculation based on the RGM has been performed.
The energies of $cccc\bar{b}$ and $bbbb\bar{c}$ systems are listed in Table \ref{table ccccb}.
For $cccc\bar{b}$ systems, the energy of each single channel is above the corresponding theoretical threshold, which means that each color singlet channel is unbound.
The energy of each hidden-color channel is much higher than that of the color singlet channel.
After the channel coupling including hidden-color channels, the energy of the $J^P=1/2^-$ state is lower than the threshold of $\Omega_{ccc} B^*_c$, the binding energy of which is $-0.9$ MeV.
For $J^P=3/2^-$ state, we first consider only the channel coupling of color singlet channels.
In this case, the calculation results show that the energy is still higher than the theoretical threshold.
However, after taking into account the channel coupling that includes the hidden-color channel, a bound state with binding energy of $-0.5$ MeV is obtained.
Besides, there is only one channel $\Omega_{ccc} B^*_c$ for the $J^P=5/2^-$ state, and this state in ubound.
These numerical results are consistent with the behavior of two clusters in the effective potential.
Both $J^P=1/2^-$ and $J^P=3/2^-$ systems exist attraction between baryon and meson, which leads to the formation of bound states.
As for $J^P=5/2^-$ systems, no bound state is formed, which is consistent with the repulsion in effective potential and Pauli exclusion principle.

Besides, from Table \ref{table ccccb}, one may note that the mass difference between $B_c$ and $B^*_c$ is too small.
This is mainly due to the parameters we used in this work.
Since some of the heavy flavor hadrons have not been observed yet, such as $B^*_c$ and $\Omega_{ccb}$ , we choose to use existing parameters instead of fitting new ones.
As we mentioned in Section II, the parameters we used here are taken from our previous work of Refs.~\cite{Huang:2020bmb, Huang:2019esu} directly to avoid adjusting the parameters.
The work of Ref.~\cite{Huang:2019esu} is about the $N\Omega_{ccc}$ and $N\Omega_{bbb}$ systems and Ref.~\cite{Huang:2020bmb} is about the fully heavy dibaryons, in which the parameters were determined by fitting the masses of both the light baryons and the heavy baryons.
For instance, the masses of the $\Omega_{ccc}$ and $\Omega_{bbb}$ are respectively $5069.1$ MeV and $15111.6$ MeV in this work.
When extending to the mesons, none of the parameters were readjusted.
So the masses of $B_c$ and $B^*_c$ are not well reproduced here, which are 6711.6 Mev and 6712.3 MeV, respectively.

In order to minimize the theoretical errors and to compare calculated results to the experimental data in future, we shift the mass of the pentaquark states here.
Generally, the mass of a pentaquark can be written as $M^{theo.}=M^{theo.}_{1}+M^{theo.}_{2}+E_{B}$, where $M^{theo.}_{1}$ and $M^{theo.}_{2}$ stand for the theoretical masses of a baryon and a meson, and $E_{B}$ is the binding energy of this state.
To minimize the theoretical errors, we can shift the mass of pentaquark to $M=M^{exp.}_{1}+M^{exp.}_{2}+E_{B}$, where the experimental
values of a baryon and a meson are used.
Since there is no experimental values of some of the fully heavy hadrons, we use the values predicted by the lattice QCD calculations~\cite{Brown:2014ena, Gregory:2009hq}, which can be seen in Table \ref{lattice hadrons}.
In Ref.~\cite{Brown:2014ena}, the masses of baryons containing one, two, or three heavy quarks were calculated systematically.
In Ref.~\cite{Gregory:2009hq}, the $B_c^*$ mass was predicted in full lattice QCD.
\begin{table}[htbp]
	\caption{\label{lattice hadrons} The masses of hadrons taken from lattice QCD, that have not yet been experimentally observed (unit: MeV).}
	\begin{tabular}{c c c|c c c} \hline\hline
		~Hadron~         & ~~~$J^P$~~~ & ~~~Mass~~~ & ~Hadron~         & ~~~$J^P$~~~ & ~~~Mass~~~ \\ \hline
        $\Omega_{ccc}$   & $3/2^+$     & 4796       & $\Omega_{bbb}$   & $3/2^+$     & 14366      \\
        $\Omega_{ccb}$   & $1/2^+$     & 8007       & $\Omega_{bbc}$   & $1/2^+$     & 11195      \\
        $\Omega_{ccb}^*$ & $3/2^+$     & 8037       & $\Omega_{bbc}^*$ & $3/2^+$     & 11229      \\
		$B_c^*$          & $ 1^-$      & 6330                                                     \\  \hline\hline
	\end{tabular}
\end{table}

Taking the pentaquark state $cccc\bar{b}$ with $J^P=1/2^-$ as an example, the calculated mass of this state is 11776.4 MeV, then the binding energy $E_B$ is obtained by subtracting the theoretical masses of $\Omega_{ccc}$ and $B_c^*$, $11776.4-5077.3-6700.0=-0.9$ (MeV).
Using the Lattice QCD masses of $\Omega_{ccc}$ and $B_c^*$, the mass of this pentaquark is $M=4796+6330+(-0.9)=11125.1$ (MeV).
So we finally obtain a fully heavy pentaquark state $cccc\bar{b}$ with $J^P=1/2^-$, the mass of which is 11125.1 MeV.
Utilizing the same operation, We can obtain the other three bound states, that are $cccc\bar{b}$ with $J^P=3/2^-$, $bbbb\bar{c}$ with $J^P=1/2^-$ and $bbbb\bar{c}$ with $J^P=3/2^-$, and the masses of these states are 11070.0 MeV, 20688.6 MeV and 20633.5 MeV, respectively.
The results of $cccc\bar{b}$ and $bbbb\bar{c}$ systems are similar, which is mainly due to the heavy flavor symmetry.


One might notice that the binding energy is much smaller than the mass of the pentaquark system. In fact, the large mass of the pentaquark system is mainly due to the use of the constituent quark masses. In the process of calculating the binding energy, we subtract the corresponding threshold energy, which also subtracts the influence of the quark mass term. Therefore, compared with other residual terms, this binding energy is not so small. We have examined the dependence of the binding energy of the system on parameters when calculating the fully heavy dibaryons, and the numerical results show that the binding energies are relatively stable~\cite{Huang:2020bmb}.

\begin{table}[htbp]
	\caption{\label{RMS ccccb} The corrected masses and RMS of $cccc\bar{b}$ and $bbbb\bar{c}$ bound states.}
	\begin{tabular}{c c c c} \hline\hline
		~~~~~~~~~~~~~~~ & ~~~$J^P$~~~ & ~~Mass (MeV)~~ & ~~RMS (fm)~~   \\ \hline
		$cccc\bar{b}$   & $1/2^-$     & 11125.1       & 1.80    \\
		$cccc\bar{b}$   & $3/2^-$     & 11070.0       & 1.87    \\
		$bbbb\bar{c}$   & $1/2^-$     & 20688.6       & 1.68    \\
		$bbbb\bar{c}$   & $3/2^-$     & 20633.5       & 1.78    \\  \hline\hline
	\end{tabular}
\end{table}

In addition, to further confirm the existence of any bound state and investigate the structure of the state, we can calculate the RMS of the states discussed above.
It is worth noting that, the scattering states have no real RMS since the relative motion wave functions of the scattered states are not integrable in the infinite space.
If we calculate the RMS of scattering states in a limited space, we can only obtain a value that increases with the expansion of computing space.
So we can calculate the RMS of various states to identify the nature of these states by keep expanding the computing space.
For the $\Omega_{ccc} B^*_c$ with $J^P=5/2^-$, the RMS in a limited space is larger than 3 fm, and it continues to increase as the computational space increases.
This indicates that the $J^P=5/2^-$ $\Omega_{ccc} B^*_c$ is not a bound state. For the $cccc\bar{b}$ and $bbbb\bar{c}$ systems with $J^P = 1/2^-$ or $3/2^-$, the value of RMS is around 1.8 fm, and it is stable as the computational space changes. This indicates that these systems are all loosely bound states. The results of RMS of the bound states are listed in the Table~\ref{RMS ccccb}.

\subsection{$cccb\bar{c}$, $cccb\bar{b}$, $bbbc\bar{c}$ and $bbbc\bar{b}$ systems}
\begin{figure*}[htbp]
	\centering
	\includegraphics[width=17cm]{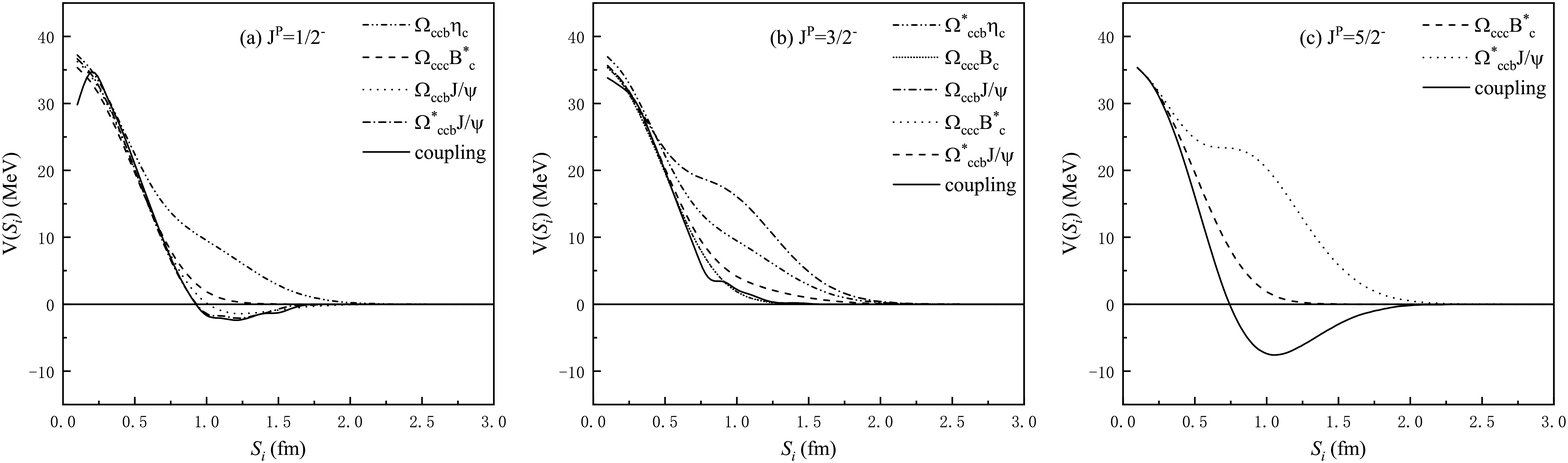}\
	\caption{\label{cccbc} The effective potentials of $cccb\bar{c}$ systems.}
\end{figure*}
The behavior of effective potential of $cccb\bar{c}$, $cccb\bar{b}$, $bbbc\bar{c}$ and $bbbc\bar{b}$ pentaquark systems is similar.
To save space, we only show the effective potential of $cccb\bar{c}$ system in Fig.~\ref{cccbc}.
One can see that, the potential of each channel is repulsive except for the one of $\Omega_{ccb} J/\psi$ and $\Omega_{ccb}^* J/\psi$ channels. However, the attraction of $\Omega_{ccb} J/\psi$ and $\Omega_{ccb}^* J/\psi$ channels is quite feeble. After channel coupling calculation, the attraction of $cccb\bar{c}$ system with $J^P=1/2^-$ is not reinforced, and there is still no obvious attraction in $cccb\bar{c}$ system $J^P=3/2^-$.
As for $cccb\bar{c}$ system of $J^P=5/2^-$, although the single channels are repulsive, attraction appears after coupling the hidden-color channel.
So it is possible for this system to form a bound state.

\begin{table*}[htbp]
	\caption{\label{table cccbc}The energies of the $cccb\bar{c}$ ,$bbbc\bar{b}$ ,$cccb\bar{b}$ and $bbbc\bar{c}$ pentaquark systems (unit: MeV).}
	\begin{tabular}{c c c c c c c c c c c c c c c} \hline\hline
		
		\multicolumn{7}{c}{$cccb\bar{c}$ systems}  &~~~~~& \multicolumn{7}{c}{$bbbc\bar{b}$ systems}  \\ \cline{1-7} \cline{9-15}
		$J^P$~ & channel & ~$E_{th}$~ & ~$E_{sc}$~  & ~~$E_{cc1}$~~ & ~~$E_{cc2}$~~& $E_B$  &~~~~&
		$J^P$~ & channel & ~$E_{th}$~ & ~$E_{sc}$~  & ~~$E_{cc1}$~~ & ~~$E_{cc2}$~~& $E_B$  \\ \cline{1-7} \cline{9-15}
		
		$\frac{1}{2}^-$ &$\Omega_{ccb} \eta_{c}$       &~11773.9~&~11775.8~&~11775.6~&~11775.4~& ub &~~~~&
		$\frac{1}{2}^-$ &$\Omega_{bbb} B^*_c$          & 21828.1 & 21828.9 & 21828.1 & 21828.0 & ub \\
		~               &$\Omega_{ccc} B^*_c$          & 11776.3 & 11778.9 & ~       & ~       &    &~~~~&
		~               &$\Omega_{bbc} \eta_{b}$       & 21827.3 & 21828.3 & ~       & ~       &    \\
		~               &$\Omega_{ccb} J/\psi$         & 11776.7 & 11778.2 & ~       & ~       &    &~~~~&
		~               &$\Omega_{bbc} \Upsilon$       & 21827.5 & 21828.3 & ~       & ~       &    \\
		~               &$\Omega^*_{ccb} J/\psi$       & 11777.3 & 11778.7 & ~       & ~       &    &~~~~&
		~               &$\Omega^*_{bbc} \Upsilon$     & 21828.1 & 21828.8 & ~       & ~       &    \\
		~               &$\Omega_{ccb8} \eta_{c8}$     & ~       & 12033.5 & ~       & ~       &    &~~~~&
		~               &$\Omega_{bbc8} \eta_{b8}$     & ~       & 22027.2 & ~       & ~       &    \\
		~               &$\Omega_{ccb8} J/\psi_{c8}$   & ~       & 12070.3 & ~       & ~       &    &~~~~&
		~               &$\Omega_{bbc8} \Upsilon_{8}$  & ~       & 22042.4 & ~       & ~       &    \\
		~               &$\Omega_{ccb8}^* J/\psi_{c8}$ & ~       & 12154.4 & ~       & ~       &    &~~~~&
		~               &$\Omega_{bbc8}^* \Upsilon_{8}$& ~       & 22141.5 & ~       & ~       &    \\
		~               &$\Omega_{ccc8} B_{c8}$        & ~       & 12106.6 & ~       & ~       &    &~~~~&
		~               &$\Omega_{bbb8} B_{c8}$        & ~       & 22099.1 & ~       & ~       &    \\
		~               &$\Omega_{ccc8} B^*_{c8}$      & ~       & 12106.7 & ~       & ~       &    &~~~~&
		~               &$\Omega_{bbb8} B^*_{c8}$      & ~       & 22099.0 & ~       & ~       &    \\ \cline{1-7} \cline{9-15}
		
		$\frac{3}{2}^-$ &$\Omega^*_{ccb} \eta_{c}$     & 11774.5 & 11776.3 & 11776.0 & 11775.7 & ub &~~~~&
		$\frac{3}{2}^-$ &$ \Omega_{bbb} B_{c}$         & 21827.3 & 21828.2 & 21828.1 & 21828.0 & ub \\
		~               &$\Omega_{ccc} B_{c}$          & 11776.5 & 11778.2 & ~       & ~       &    &~~~~&
		~               &$\Omega_{bbb} B^*_{c}$        & 21828.1 & 21828.9 & ~       & ~       &    \\
		~               &$\Omega_{ccb} J/\psi$         & 11776.7 & 11778.7 & ~       & ~       &    &~~~~&
		~               &$\Omega_{bbc} \Upsilon$       & 21827.5 & 21828.6 & ~       & ~       &    \\
		~               &$\Omega_{ccc} B^*_{c}$        & 11777.3 & 11778.9 & ~       & ~       &    &~~~~&
		~               &$\Omega^*_{bbc} \eta_{b}$     & 21827.9 & 21828.9 & ~       & ~       &    \\
		~               &$\Omega^*_{ccb} J/\psi$       & 11777.3 & 11779.0 & ~       & ~       &    &~~~~&
		~               &$\Omega^*_{bbc} \Upsilon$     & 21828.1 & 21829.0 & ~       & ~       &    \\
		~               &$\Omega_{ccb8} J/\psi_{8}$    & ~       & 12006.5 & ~       & ~       &    &~~~~&
		~               &$\Omega_{bbc8} \Upsilon_{8}$  & ~       & 22004.5 & ~       & ~       &    \\
		~               &$\Omega_{ccc8} B^*_{c8}$      & ~       & 12106.5 & ~       & ~       &    &~~~~&
		~               &$\Omega_{bbb8} B^*_{c8}$      & ~       & 22099.1 & ~       & ~       &    \\
		~               &$\Omega_{ccb8}^* \eta_{c8}$   & ~       & 11948.0 & ~       & ~       &    &~~~~&
		~               &$\Omega_{bbc8}^* \eta_{b8}$   & ~       & 21963.8 & ~       & ~       &    \\
		~               &$\Omega_{ccb8}^* J/\psi_{c8}$ & ~       & 12080.9 & ~       & ~       &    &~~~~&
		~               &$\Omega_{bbc8}^* \Upsilon_{8}$& ~       & 22077.0 & ~       & ~       &    \\ \cline{1-7} \cline{9-15}
		
		$\frac{5}{2}^-$ & $\Omega_{ccc} B^*_c$         & 11777.3 & 11778.9 & 11778.4 & 11776.3 & $-1.0$ &~~~~&
		$\frac{5}{2}^-$ & $\Omega_{bbb} B^*_c$         & 21828.1 & 21828.9 & 21828.7 & 21820.6 & $-7.5$\\
		~               & $\Omega_{ccb}^* J/\psi$      & 11777.3 & 11779.3 & ~       & ~       &     &~~~~&
		~               &$\Omega^*_{bbc} \Upsilon$     & 21828.1 & 21829.1 &         & ~       &     \\
		~               & $\Omega_{ccb8}^* J/\psi_{c8}$& ~       & 11948.0 & ~       & ~       &     &~~~~&
		~               &$\Omega^*_{bbc8} \Upsilon_{8}$&         & 21963.8 &         & ~       &     \\ \hline\hline

		\multicolumn{7}{c}{$cccb\bar{b}$ systems}  &~~~~~& \multicolumn{7}{c}{$bbbc\bar{c}$ systems}  \\ \cline{1-7} \cline{9-15}
        $J^P$~ & channel & ~$E_{th}$~ & ~$E_{sc}$~  & ~~$E_{cc1}$~~ & ~~$E_{cc2}$~~& $E_B$  &~~~~&
        $J^P$~ & channel & ~$E_{th}$~ & ~$E_{sc}$~  & ~~$E_{cc1}$~~ & ~~$E_{cc2}$~~& $E_B$  \\ \cline{1-7} \cline{9-15}
		
		$\frac{1}{2}^-$ &$\Omega_{ccb} B_c$            &~15125.6~&~15127.2~&~15127.0~&~15126.2~& ub &~~~~&
		$\frac{1}{2}^-$ &$\Omega_{bbb} J/\psi$         & 18478.4 & 18479.5 & 18477.0 & 18476.3 & ub \\
		~               &$\Omega_{ccb} B^*_c$          & 15126.4 & 15127.7 & ~       & ~       &    &~~~~&
		~               &$\Omega_{bbc} B_{c}$          & 18475.9 & 18477.2 & ~       & ~       &    \\
		~               &$\Omega^*_{ccb} B^*_c$        & 15126.9 & 15128.1 & ~       & ~       &    &~~~~&
		~               &$\Omega_{bbc} B^*_{c}$        & 18476.6 & 18477.7 & ~       & ~       &    \\
		~               &$\Omega_{ccc}\Upsilon$        & 15128.2 & 15129.5 & ~       & ~       &    &~~~~&
		~               &$\Omega^*_{bbc}B^*_{c}$       & 18477.2 & 18478.1 & ~       & ~       &    \\
		~               &$\Omega_{ccb8} B_{c8}$        & ~       & 15359.0 & ~       & ~       &    &~~~~&
		~               &$\Omega_{bbc8}  B_{c8}$       & ~       & 18702.5 & ~       & ~       &    \\
		~               &$\Omega_{ccb8} B^*_{c8}$      & ~       & 15376.4 & ~       & ~       &    &~~~~&
		~               &$\Omega_{bbc8} B^*_{c8}$      & ~       & 18717.3 & ~       & ~       &    \\
		~               &$\Omega_{ccb8}^* B^*_{c8}$    & ~       & 15483.0 & ~       & ~       &    &~~~~&
		~               &$\Omega_{bbc8}^* B^*_{c8}$    & ~       & 18870.7 & ~       & ~       &    \\
		~               &$\Omega_{ccc8} \eta_{b8}$     & ~       & 15433.3 & ~       & ~       &    &~~~~&
		~               &$\Omega_{bbb8} \eta_{c8}$     & ~       & 18773.1 & ~       & ~       &    \\
		~               &$\Omega_{ccc8} \Upsilon_{8}$  & ~       & 15433.3 & ~       & ~       &    &~~~~&
		~               &$\Omega_{bbb8}J/\psi_{8}$     & ~       & 18772.8 & ~       & ~       &    \\ \hline
		
		$\frac{3}{2}^-$ &$\Omega_{ccc} \eta_{b}$       & 15128.0 & 15129.3 & 15127.4 & 15127.3 & ub &~~~~&
		$\frac{3}{2}^-$ &$\Omega_{bbb} \eta_{c}$       & 18475.6 & 18476.8 & 18476.6 & 18476.2 & ub \\
		~               &$\Omega_{ccc} \Upsilon$       & 15128.2 & 15129.5 & ~       & ~       &    &~~~~&
		~               &$\Omega_{bbb} J/\psi$         & 18478.4 & 18479.5 & ~       & ~       &    \\
		~               &$\Omega_{ccb} B^*_c$          & 15126.4 & 15128.0 & ~       & ~       &    &~~~~&
		~               &$\Omega_{bbc} B^*_c$          & 18476.6 & 18478.0 & ~       & ~       &    \\
		~               &$\Omega^*_{ccb} B_c$          & 15126.2 & 15127.8 & ~       & ~       &    &~~~~&
		~               &$\Omega^*_{bbc} B_c$          & 18476.5 & 18477.7 & ~       & ~       &    \\
		~               &$\Omega^*_{ccb} B^*_c$        & 15126.9 & 15128.4 & ~       & ~       &    &~~~~&
		~               &$\Omega^*_{bbc}B^*_c$         & 18477.2 & 18478.4 & ~       & ~       &    \\
		~               &$\Omega_{ccb8} B^*_{c8}$      & ~       & 15331.9 & ~       & ~       &    &~~~~&
		~               &$\Omega_{bbc8} B^*_{c8}$      & ~       & 18679.9 & ~       & ~       &    \\
		~               &$\Omega_{ccc8} \Upsilon_{8}$  & ~       & 15433.3 & ~       & ~       &    &~~~~&
		~               &$\Omega_{bbb8} J/\psi_{8}$    & ~       & 18772.8 & ~       & ~       &    \\
		~               &$\Omega_{ccb8}^* B_{c8}$      & ~       & 15267.9 & ~       & ~       &    &~~~~&
		~               &$\Omega_{bbc8}^* B_{c8}$      & ~       & 18639.9 & ~       & ~       &    \\
		~               &$\Omega_{ccb8}^* B^*_{c8}$    & ~       & 15407.6 & ~       & ~       &    &~~~~&
		~               &$\Omega_{bbc8}^* B^*_{c8}$    & ~       & 18751.4 & ~       & ~       &    \\ \hline
		
		$\frac{5}{2}^-$ &$\Omega_{ccc} \Upsilon$       & 15128.2 & 15129.5 & 15128.2 & 15125.0 & $-1.9$ &~~~~&
		$\frac{5}{2}^-$ &$\Omega_{bbb} J/\psi$         & 18478.4 & 18479.5 & 18478.2 & 18473.5 & $-3.7$ \\
		~               &$\Omega_{ccb} B^*_c$          & 15126.9 & 15128.6 & ~       & ~       &    &~~~~&
		~               &$\Omega^*_{bbc} B^*_c$        & 18477.2 & 18478.6 & ~       & ~       &    \\
		~               &$\Omega_{ccb8} B^*_{c8}$      & ~       & 15273.0 & ~       & ~       &    &~~~~&
		~               &$\Omega^*_{bbc8} B^*_{c8}$    & ~       & 18639.9 & ~       & ~       &    \\ \hline\hline

	\end{tabular}
\end{table*}

The energies of the $cccb\bar{c}$ ,$bbbc\bar{b}$ ,$cccb\bar{b}$ and $bbbc\bar{c}$ pentaquark systems are presented in Table \ref{table cccbc}.
On the basis of the numerical results, all the single channels and channel coupling of color singlet channels are unbound.
After coupling hidden-color channels, energies of systems with $J^P=1/2^-$ and $J^P=3/2^-$ remain above the corresponding thresholds.
In regard to systems with $J^P=5/2^-$, unlike $cccc\bar{b}$ and $bbbb\bar{c}$ systems, the Pauli principle doesn't rule out the possibility of bound states here.
Four bound states are formed after channel coupling, which are $cccb\bar{c}$ system with binding energy of $-1.0$ MeV, $bbbc\bar{b}$ system with binding energy of $-7.3$ MeV, $cccb\bar{b}$ system with binding energy of $-1.9$ MeV and $bbbc\bar{c}$ system with binding energy of $-3.7$ MeV.

\begin{table}[htbp]
	\caption{\label{RMS cccbc} The corrected masses and RMS of $cccb\bar{c}$, $bbbc\bar{b}$, $cccb\bar{b}$ and $bbbc\bar{c}$ bound states.}
	\begin{tabular}{c c c c} \hline\hline
		~~~~~~~~~~~~~~~ & ~~~$J^P$~~~ & ~~Mass (MeV)~~ & ~~RMS (fm)~~   \\ \hline
		$cccb\bar{c}$   & $5/2^-$     & 11125.0       & 1.84    \\
		$bbbc\bar{b}$   & $5/2^-$     & 20681.8       & 1.73    \\
		$cccb\bar{b}$   & $5/2^-$     & 14254.4       & 1.79    \\
		$bbbc\bar{c}$   & $5/2^-$     & 17459.2       & 1.73    \\  \hline\hline
	\end{tabular}
\end{table}

Same as previous sector, to minimize the theoretical errors and make sense of the results, we perform the mass correction here.
Meanwhile, the RMS of the bound states are also calculated, which are listed in the Table~\ref{RMS cccbc}.
According to the RMS results, the spatial structure of these four bound states are not compact.
In other words, high-spin $cccb\bar{c}$ ,$bbbc\bar{b}$ ,$cccb\bar{b}$ and $bbbc\bar{c}$ pentaquark states are loosely bound states in our calculations.

\subsection{$ccbb\bar{c}$ and $ccbb\bar{b}$ systems}
\begin{figure*}[htbp]
	\centering
	\includegraphics[width=17cm]{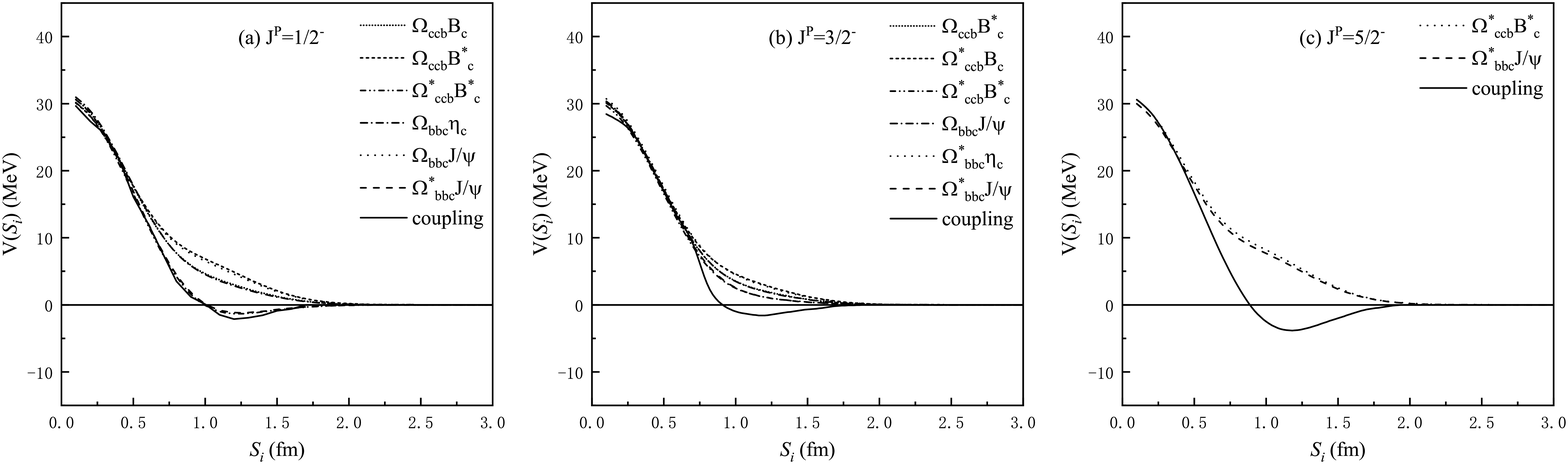}\
	\caption{\label{ccbbc} The effective potentials of $ccbb\bar{c}$ systems.}
\end{figure*}

In this sector, we investigate the $ccbb\bar{c}$ and $ccbb\bar{b}$ systems with quantum numbers of $J^P=1/2^-$, $J^P=3/2^-$ and $J^P=5/2^-$.
The effective potential of $ccbb\bar{c}$ is shown in the Fig. \ref{ccbbc}.
As one can see that the effective potentials of both $\Omega_{ccb}^* B_c^*$ and $\Omega_{bbc}^* J/\psi$ channels are slightly attractive, while those of other single channels are all repulsive.
After the channel coupling calculation, the attraction generated in the three $J^P=1/2^-$, $J^P=3/2^-$ and $J^P=5/2^-$ systems is still very weak.
Based on the bound-state calculation, we find that all the color-singlet channels are unbound, and the energies of all hidden-color channels are much higher than the theoretical thresholds. In addition, no bound state is formed in either color singlet channel coupling or channel coupling including hidden-color channels.
The energies of both $ccbb\bar{c}$ and $ccbb\bar{b}$ systems with different quantum numbers are listed in the Table ~\ref{table ccbbc}.

\begin{table*}[htbp]
	\caption{\label{table ccbbc}The energies of the $ccbb\bar{c}$ and $ccbb\bar{c}$ pentaquark systems (unit: MeV).}
	\begin{tabular}{c c c c c c c c c c c c c c c} \hline\hline
		
		\multicolumn{7}{c}{$ccbb\bar{c}$ systems}  &~~~~~& \multicolumn{7}{c}{$ccbb\bar{b}$ systems}  \\ \cline{1-7} \cline{9-15}
        $J^P$~ & channel & ~$E_{th}$~ & ~$E_{sc}$~  & ~~$E_{cc1}$~~ & ~~$E_{cc2}$~~& $E_B$  &~~~~&
        $J^P$~ & channel & ~$E_{th}$~ & ~$E_{sc}$~  & ~~$E_{cc1}$~~ & ~~$E_{cc2}$~~& $E_B$  \\ \cline{1-7} \cline{9-15}
		
		$\frac{1}{2}^-$ &$\Omega_{ccb}B_c$          &~15125.6~&~15127.1~&~15125.6~&~15125.5~& ub   &~~~~&
		$\frac{1}{2}^-$ &$\Omega_{ccb}\eta_{b}$     & 18477.1 & 18478.3 & 18477.0 & 19476.9 & ub   \\
		~               &$\Omega_{ccb} B^*_c$       & 15126.4 & 15127.9 & ~       & ~       &      &~~~~&
		~               &$\Omega_{ccb} \Upsilon$    & 18477.3 & 18478.5 & ~       & ~       &      \\
		~               &$\Omega^*_{ccb} B^*_c$     & 15126.9 & 15128.2 & ~       & ~       &      &~~~~&
		~               &$\Omega^*_{ccb} \Upsilon$  & 18477.8 & 18478.8 & ~       & ~       &      \\
		~               &$\Omega_{bbc} \eta_{c}$    & 15124.2 & 15128.2 & ~       & ~       &      &~~~~&
		~               &$\Omega_{bbc} B_c$         & 18475.9 & 18477.1 & ~       & ~       &      \\
		~               &$\Omega_{bbc} J/\psi$      & 15127.0 & 15128.5 & ~       & ~       &      &~~~~&
		~               &$\Omega_{bbc} B^*_{c}$     & 18476.6 & 18477.9 & ~       & ~       &      \\
		~               &$\Omega^*_{bbc} J/\psi$    & 15127.6 & 15128.8 & ~       & ~       &      &~~~~&
		~               &$\Omega^*_{bbc} B^*_{c}$   & 18477.2 & 18477.9 & ~       & ~       &      \\
		~               &$\Omega_{ccb8} B_{c8}$     & ~       & 15400.8 & ~       & ~       &      &~~~~&
		~               &$\Omega_{ccb8} \eta_{b8}$  & ~       & 18725.4 & ~       & ~       &      \\	
		~               &$\Omega_{ccb8} B^*_{c8}$   & ~       & 15422.8 & ~       & ~       &      &~~~~&
		~               &$\Omega_{ccb8}\Upsilon_{8}$& ~       & 18745.3 & ~       & ~       &      \\
		~               &$\Omega_{ccb8}^* B^*_{c8}$ & ~       & 15452.7 & ~       & ~       &      &~~~~&
		~             &$\Omega_{ccb8}^*\Upsilon_{8}$& ~       & 18780.0 & ~       & ~       &      \\
		~               &$\Omega_{bbc8} \eta_{c8}$  & ~       & 15398.2 & ~       & ~       &      &~~~~&
		~               &$\Omega_{bbc8} B_{c8}$     & ~       & 18722.3 & ~       & ~       &      \\
		~               &$\Omega_{bbc8} J/\psi_{8}$ & ~       & 15425.1 & ~       & ~       &      &~~~~&
		~               &$\Omega_{bbc8} B^*_{c8}$   & ~       & 18747.9 & ~       & ~       &      \\
		~               &$\Omega_{bbc8}^*J/\psi_{8}$& ~       & 15471.8 & ~       & ~       &      &~~~~&
        ~               &$\Omega_{bbc8}^* B^*_{c8}$ & ~       & 18799.2 & ~       & ~       &      \\  \hline
		
	    $\frac{3}{2}^-$ &$\Omega_{ccb} B^*_{c}$     &~15126.4~&~15127.8~&~15126.2~&~15126.0~& ub   &~~~~&
	    $\frac{3}{2}^-$&$\Omega_{ccb} \Upsilon$     & 18477.3 & 18478.4 & 18477.5 & 18477.5 & ub   \\
		~               &$\Omega^*_{ccb} B_{c}$     & 15126.2 & 15127.7 & ~       & ~       &      &~~~~&
		~               &$\Omega^*_{ccb} \eta_{b}$  & 18477.6 & 18478.8 & ~       & ~       &      \\
		~               &$\Omega^*_{ccb} B^*_c$     & 14126.9 & 15128.3 & ~       & ~       &      &~~~~&
		~               &$\Omega^*_{ccb} \Upsilon$  & 18477.8 & 18478.9 & ~       & ~       &      \\
		~               &$\Omega_{bbc} J/\psi$      & 15127.0 & 15128.4 & ~       & ~       &      &~~~~&
		~               &$\Omega_{bbc} B^*_{c}$     & 18476.6 & 18477.8 & ~       & ~       &      \\
		~               &$\Omega^*_{bbc} \eta_{c}$  & 15124.8 & 15126.2 & ~       & ~       &      &~~~~&
		~               &$\Omega^*_{bbc}B_{c}$      & 18476.5 & 18477.6 & ~       & ~       &      \\
		~               &$\Omega^*_{bbc}J/\psi$     & 15127.6 & 15128.9 & ~       & ~       &      &~~~~&
		~               &$\Omega^*_{bbc}B^*_{c}$    & 18477.2 & 18478.3 & ~       & ~       &      \\
		~               &$\Omega_{ccb8} B^*_{c8}$   & ~       & 15389.5 & ~       & ~       &      &~~~~&
		~               &$\Omega_{ccb8}\Upsilon_{8}$& ~       & 18713.9 & ~       & ~       &      \\
		~               &$\Omega_{ccb8}^* B_{c8}$   & ~       & 15400.9 & ~       & ~       &      &~~~~&
		~               &$\Omega_{ccb8}^*\eta_{b8}$ & ~       & 18726.7 & ~       & ~       &      \\
		~               &$\Omega_{ccb8}^* B^*_{c8}$ & ~       & 15405.3 & ~       & ~       &      &~~~~&
		~             &$\Omega_{ccb8}^*\Upsilon_{8}$& ~       & 18717.0 & ~       & ~       &      \\
		~               &$\Omega_{bbc8} J/\psi_{8}$ & ~       & 15384.0 & ~       & ~       &      &~~~~&
		~               &$\Omega_{bbc8} B^*_{c8}$   & ~       & 18708.7 & ~       & ~       &      \\
		~               &$\Omega_{bbc8}^* \eta_{c8}$& ~       & 15408.0 & ~       & ~       &      &~~~~&
		~               &$\Omega_{bbc8}^* B_{c8}$   & ~       & 18733.7 & ~       & ~       &      \\
		~               &$\Omega_{bbc8}^*J/\psi_{8}$& ~       & 15412.3 & ~       & ~       &      &~~~~&
		~               &$\Omega_{bbc8}^* B_{c8}^*$ & ~       & 18724.1 & ~       & ~       &      \\ \hline
		
		$\frac{5}{2}^-$ & $\Omega_{ccb} B^*_c$      & 15126.9 & 15128.5 & 15128.5 & 15128.1 & ub   &~~~~&
		$\frac{5}{2}^-$ & $\Omega_{ccb} \Upsilon$   & 18477.8 & 18479.1 & 18478.3 & 18478.1 & ub   \\
		~               &$\Omega_{bbc} J/\psi$      & 15127.6 & 15129.1 & ~       & ~       &      &~~~~&
		~               &$\Omega_{bbc} B^*_c$       & 18477.2 & 18478.4 & ~       & ~       &      \\
		~               &$\Omega_{ccb8} B^*_{c8}$   & ~       & 15362.3 & ~       & ~       &      &~~~~&
		~               &$\Omega_{ccb8}\Upsilon_{8}$& ~       & 18686.3 & ~       & ~       &      \\
		~               &$\Omega_{bbc8} J/\psi_{8}$ & ~       & 15361.5 & ~       & ~       &      &~~~~&
		~               &$\Omega_{bbc8} B^*_{c8}$   & ~       & 18684.9 & ~       & ~       &      \\	\hline\hline
		
	\end{tabular}
\end{table*}

\section{Summary}
In this work, we systematically investigate the low-lying fully heavy pentaquark systems in the chiral quark model.
First, an adiabatic calculation of the effective potential is performed to explore the interaction between the baryon and meson clusters.
Both the single channel and the channel coupling calculation are performed to explore the effect of the multi-channel coupling.
Two types of color structure $\bf{1}$$ _{[QQQ]} \bf{\otimes} \bf{1}$$_{[Q \bar{Q}]}$ and $\bf{8}$$ _{[QQQ]} \bf{\otimes} \bf{8}$$_{[Q \bar{Q}]}$ are taken into account.
The dynamic bound-state calculation is carried out to search for any bound state in the fully heavy pentaquark systems.
Meanwhile, the RMS of the fully heavy pentaquark states is calculated to confirm the existence of any bound state and explore the internal spatial structure of the bound states.

The numerical results show that the effect of the channel coupling with hidden-color channels is important for forming bound states of the fully heavy pentaquark systems.
With the help of the channel-coupling, we obtain eight fully heavy pentaquark states, which are $cccc\bar{b}$ with $J^P = 1/2^-$ and mass of 11125.1 MeV, $cccc\bar{b}$ with $J^P = 3/2^-$ and mass of 11070.0 MeV, $bbbb\bar{c}$ with $J^P = 1/2^-$ and mass of 20688.6 MeV, $bbbb\bar{c}$ with $J^P = 3/2^-$ and mass of 20633.5 MeV, $cccb\bar{c}$ with $J^P = 5/2^-$ and mass of 11125.0 MeV, $bbbc\bar{b}$ with $J^P = 5/2^-$ and mass of 20681.8 MeV, $cccb\bar{b}$ with $J^P = 5/2^-$ and mass of 14254.4 MeV and $bbbc\bar{c}$ with $J^P = 5/2^-$ and mass of 17459.2 MeV.
Since the binding energies of these states are all below 10 MeV and the RMS are around 1.8 fm, we tend to interpret these states as molecular states, which are worth looking for in future experiments.

Encouraged by the discovery of the fully charmed tetraquark states in experiments, some theoretical work has begun to search for the existence of fully heavy pentaquark states.
By using the QCD sum rule approach, both the fully charm pentaquark and fully bottom pentaquark candidates are obtained~\cite{Zhang:2020vpz,Wang:2021xao}. The calculated mass is $7.41^{+0.27}_{-0.31}$ GeV for $cccc\bar{c}$ and $21.60^{+0.73}_{-0.22}$ GeV for $bbbb\bar{b}$ in Ref.~\cite{Zhang:2020vpz}, while it is $7.93\pm 0.15$ GeV for $cccc\bar{c}$ and $23.91\pm 0.15$ GeV for $bbbb\bar{b}$ in Ref.~\cite{Wang:2021xao}.
In the chiral quark model and quark delocalization color screening model, the $cccc\bar{c}$ state is obtained with $J^P = 1/2^-$ and the mass of $7891.9 \sim 7892.7$ MeV, the $bbbb\bar{b}$ state is obtained with $J^P=1/2^-$ and the mass of $23810.1 \sim 23813.8$ MeV, and with $J^P = 3/2^-$ and the mass of $23748.2 \sim 23752.3$ MeV~\cite{Yan:2021glh}.
In a lattice-QCD inspired quark model, several resonances were obtained in each spin-parity channel for the fully charm and bottom systems, with mass above $8$ GeV and $24$ GeV, respectively~\cite{Yang:2022bfu}.
In the framework of CMI model, two stable candidates were obtained, which are $J^P = 3/2^-$ $ccbb\bar{b}$ with mass of 17416 MeV and $J^P = 5/2^-$ $ccbb\bar{b}$ with mass of 17477 MeV~\cite{An:2020jix}.
However, in the work of Ref.~\cite{An:2022fvs}, no any stable fully heavy pentaquark state was found within the constituent quark model.
Therefore, various theoretical approaches may lead to different conclusions. We hope to have more theoretical and experimental work to search for possible fully heavy pentaquark states, and further test and improve theoretical methods.

\acknowledgments{This work is supported partly by the National Science Foundation
of China under Contract Nos. 11675080, 11775118 and 11535005.}

\end{document}